\documentclass[twocolumn,english,pre,floatfix,citeautoscript,nofootinbib]{revtex4}
\usepackage{amsbsy}
\usepackage{latexsym,epsfig,graphicx}
\usepackage{dcolumn}
\usepackage{subfigure}
\usepackage{comment}
\usepackage{color}
\usepackage[colorlinks,urlcolor=blue,citecolor=blue]{hyperref}
\usepackage{amstext}
\usepackage{amssymb}
\usepackage{setspace}
\usepackage{amsmath}
\usepackage{makeidx}
\usepackage{bm}
\usepackage{ulem}
\usepackage{multirow}
\usepackage{mathrsfs}

\usepackage{array}

\begin{document}
\title{Controllable excitation of vector Akhmediev breather patterns}
\author{Yan-Hong Qin$^{1}$}\email{yhqin@xju.edu.cn}
\author{Ning Cao$^{2}$}
\author{Li-Chen Zhao$^{2,3,4,5}$}\email{zhaolichen3@nwu.edu.cn}
\address{$^{1}$School of Physical Science and Technology, Xinjiang University, Urumqi, 830046, China}
\address{$^{2}$School of Physics, Northwest University, Xi'an 710127, China}
\address{$^{3}$Shaanxi Key Laboratory for Theoretical Physics Frontiers, Xi'an 710127, China}
\address{$^{4}$Peng Huanwu Center for Fundamental Theory, Xi'an 710127, China}
\address{$^{5}$Fundamental Discipline Research Center for  Quantum Science and technology of Shaanxi Province, Xi'an 710127, China}
\begin{abstract}
In the focusing Manakov system, multiple modulation instability (MI) branches coexist on the same plane wave background, so the usual weak periodic modulation cannot selectively excite a single vector Akhmediev breather (AB). Here we propose an eigenvector-based initial perturbation scheme that constructs the initial condition as a plane wave plus Fourier modes whose coefficients follow the perturbation eigenvector of a selected MI branch, enabling controllable high-fidelity excitation of desired vector ABs. Numerical simulations show near-100\% fidelity with the exact AB solution. The underlying mechanism is eigenvector-controlled mode selection. The perturbation eigenvectors of the non-Hermitian Bogoliubov-de Gennes operator resolve coexisting MI branches at the identical perturbation frequency, enabling selection of the target MI branch. High-fidelity excitation of the prescribed AB pattern is attained either for comparable gains or when the targeted branch exhibits a sufficient gain advantage. The proposed method provides a simple and robust framework for controllable generation of vector ABs over a broad parameter range, highlighting the key role of eigenvector selectivity in multi-component nonlinear systems.

\end{abstract}

\date{\today}	
\maketitle

\section{Introduction}
	
The experimental observation of exact nonlinear wave solutions often faces a fundamental difficulty that the ideal initial conditions required by exact solutions are typically too intricate to be prepared in the experiment. Instead, one must rely on simple, experimentally accessible initial states, such as weak modulations on a plane wave background, and hope that the resulting dynamics closely match the exact solution. A classic example is the scalar AB \cite{AB1}. Although its exact initial profile is a localized, analytically constructed waveform, it can be excited to high fidelity simply by imposing a weak periodic modulation $\cos(k\tau)$ on a plane wave \cite{AB2,AB3,AB4,AB5}. The MI seeded by this perturbation selectively amplifies the corresponding Fourier mode, and the emergent breather dynamics agree remarkably well with the exact solution. This fortunate coincidence, however, is not guaranteed in more complex systems. For rogue wave solutions, for instance, experiments typically prepare initial conditions that are themselves close to the exact solution \cite{RW1,RW2,RW3}, underscoring that the use of genuinely simple initial states to observe exact solutions is, in general, a nontrivial challenge \cite{RW4,RW5}.

In multi-component vector nonlinear systems, the dynamics become far richer and more complex than in the scalar case. The Manakov model serves as the canonical two-component representative of such systems \cite{Forest}. For the defocusing two-component case, a plane wave background supports only one MI branch alongside a modulationally stable branch \cite{Baronio1,Baronio2}; consequently, the excitation behavior can be understood essentially in the same way as in the scalar case. The focusing two-component case, however, supports two distinct MI branches on the identical background \cite{Forest,lingzhao1,Qin1,lingzhao2}. Within the parameter region where these branches coexist, a weak periodic modulation can simultaneously trigger the exponential growth of multiple unstable modes \cite{lingzhao2,Liu3,Liu5}. Each mode develops into a different AB pattern, and their nonlinear competition completely erases any fidelity to a single exact vector AB solution. A clean excitation of an AB, namely the selective excitation of a prescribed MI branch that yields a well-defined vector AB pattern with high fidelity to the corresponding exact solution, can only be achieved outside the coexistence region \cite{Liu1}, i.e., in a single MI branch domain. Extending to higher components, the dynamical behavior is necessarily even richer than in the two-component case. Thus, the same simple perturbation that works efficiently in the scalar case fails in vector systems precisely because the multiplicity of unstable modes inevitably leads to an uncontrollable mixture of patterns. Realizing a predetermined vector AB pattern in a clean and controlled way remains a challenging task bridging integrable nonlinear physics and experimental physical systems.

In this work, we address the controllable excitation of vector AB patterns using simple initial perturbations that are readily realizable in experiments, even when multiple MI branches coexist. We demonstrate that one can selectively seed a single unstable mode by properly choosing the eigenvector of the initial perturbation, rather than just its frequency. The fidelity between the resulting nonlinear evolution and the exact vector AB solution can approach 100\%. More fundamentally, the non-Hermitian Bogoliubov-de Gennes (BdG) eigenvectors distinguish the coexisting MI branches at the same perturbation frequency. Constructing the initial perturbation from the right eigenvector of a selected MI branch concentrates the initial modal weight in that branch, with this branch dominating throughout the entire breather evolution. This control succeeds for comparable MI gains or a sufficiently stronger gain in the targeted branch. If the targeted branch’s gain is substantially lower, the higher-gain competing branch can develop to dominate and prevent the desired breather excitation during evolution. Our results generalize the scalar paradigm of automatic mode selection by frequency to a vector principle of deliberate eigenvector control, bridging ideal exact solutions with experimentally accessible initial states. The evolutionary characteristics of non-Hermitian systems fundamentally couple mode competition to initial excitation conditions, making the choice of eigenvector key to the controllability of multi-component integrable systems.

The remainder of this paper is organized as follows. In Sec.~\ref{sec2}, we demonstrate controllable excitation of vector ABs in the focusing Manakov system using the eigenvector-based initial perturbation scheme, achieving fidelity near 100\% with the exact solutions. In Sec.~\ref{sec3}, we analyze the physical mechanism underlying this controllability via a full BdG modal decomposition, tracking the evolution of contributions of all perturbation eigenvectors. Finally, conclusions and discussion are given in Sec.~\ref{sec4}.

\section{Perfectly Controllable Excitations of Vector Akhmediev Breathers}\label{sec2}

For multi-component vector nonlinear systems, the two-component Manakov system provides a typical representative \cite{Manakov,BdG3,fiber1,fiber2,fiber3,fiber4,fiber5,BEC1,BEC2}. It arises in a single-mode fiber when the two polarization components are coupled, forming a two-component vector field whose slowly varying envelopes $\psi_1$ and $\psi_2$ evolve according to the Manakov system. In dimensionless form, they read \cite{Manakov}:
\begin{subequations}\label{model}
\begin{align}
\mathrm{i}\frac{\partial\psi_1}{\partial\xi}+\frac{1}{2}\frac{\partial^2\psi_1}{\partial\tau^2}+\sigma(|\psi_1|^2+|\psi_2|^2)\psi_1=0,\\
\mathrm{i}\frac{\partial\psi_2}{\partial\xi}+\frac{1}{2}\frac{\partial^2\psi_2}{\partial\tau^2}+\sigma(|\psi_1|^2+|\psi_2|^2)\psi_2=0.
\end{align}
\end{subequations}
Here $\tau$ and $\xi$ denote the retarded time and the normalized propagation distance along the optical fiber, respectively. The nonlinear coefficient $\sigma=1$ and  $\sigma=-1$ correspond to the focusing and defocusing regimes, respectively \cite{Manakov,BdG3,fiber1,fiber2,fiber3,fiber4}.

Exact vector AB solutions of the system \eqref{model} have been systematically studied \cite{lingzhao2,Liu1,AB6}. Their analytical forms provide important benchmarks for the nonlinear patterns that may emerge from an unstable background. However, one cannot prepare the exact initial profile prescribed by the exact solution; instead, one must design simple and experimentally feasible initial conditions that evolve with high fidelity into a chosen breather pattern. To realize a controllable excitation of vector ABs, one must first understand what fundamentally distinguishes the vector system from its scalar counterpart. In the scalar nonlinear Schr\"{o}dinger equation, a weak modulation $\cos(k\tau)$ on a plane wave cleanly evolves into the well-known AB \cite{AB1}. The underlying MI supports only a single unstable branch at any given frequency, so the initial perturbation has no choice but to seed that specific mode, and the selectivity is naturally preserved throughout the nonlinear evolution \cite{AB2,AB3,AB4}. In the Manakov system, particularly in the focusing regime, however, the same plane wave background can simultaneously host multiple distinct MI branches at the identical frequency \cite{lingzhao1,lingzhao2,Liu1}, each with its own perturbation eigenvector and gain rate. A scalar-like modulation that ignores this eigenvector structure inevitably excites multiple modes at once, triggering a competition that destroys fidelity to any single AB pattern, and the resulting breather becomes an uncontrolled superposition of patterns. Controllable vector AB excitation therefore requires that the initial modulation be confined to a single MI eigenmode.

\begin{figure}[t!]
\centering		
\includegraphics[width=85mm]{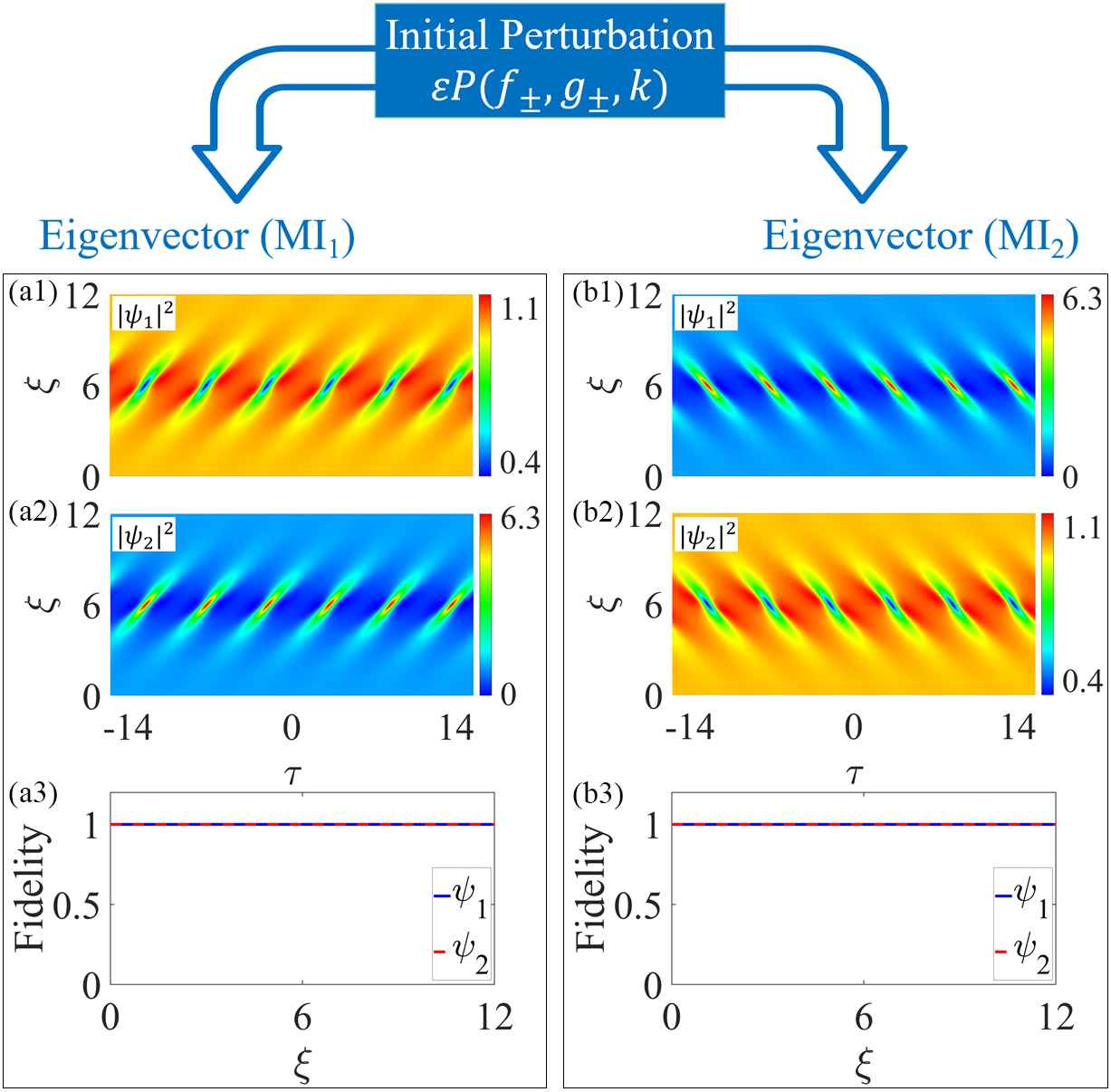}
\caption{The perfectly controllable excitations of vector ABs in the focusing Manakov system under the initial condition given by Eq.~\eqref{initial}. The parameters are $a_1=1,a_2=1,b_1=-1.2,b_2=-b_1,k=1.2$. The left plane corresponds to the MI$_1$ branch, with
 $\varepsilon= -0.00159882-0.00507872\rm{i}$; the right plane corresponds to the MI$_2$ branch, with $\varepsilon=0.00472253 + 0.0024594\rm {i}$. It is shown that the fidelity between the numerical results and the exact solutions remains nearly equal to 100\%, even when both MI branches coexist on the same background.}\label{fig1}
\end{figure}

Guided by this insight, we propose a scheme that accomplishes this by using the perturbation eigenvector for the initial perturbation with Fourier modes (see Fig.~\ref{fig1}). The initial states are then set as
\begin{subequations}\label{initial}
\begin{align}
&&\psi_1(\tau,0)=a_1e^{\mathrm{i} b_1 \tau}(1+f_+e^{\mathrm{i} k \tau}+f_-^{*}e^{-\mathrm{i} k \tau}),\\
&&\psi_2(\tau,0)=a_2e^{\mathrm{i} b_2 \tau}(1+g_+e^{\mathrm{i} k \tau}+g_-^{*}e^{-\mathrm{i} k \tau}),
\end{align}
\end{subequations}
where $a_i$ and $b_i$ denote the amplitude and frequency of the plane wave background, respectively, and $k$ is the perturbation frequency. The coefficients $f_{\pm}$ and $g_{\pm}$ are chosen according to the perturbation eigenvectors obtained from the standard linear MI analysis \cite{Baronio1,lingzhao1,lingzhao2,Qin1}.  $f_-^*$ and $g_-^*$ are the conjugate counterparts of $f_-$ and $g_-$. We substitute the perturbed plane-wave ansatz $\psi_1(\tau,\xi)=a_1e^{\theta_1}\bigl(1+p_1(\tau,\xi)\bigr)$ and $\psi_2(\tau,\xi)=a_2e^{\theta_2}\bigl(1+p_2(\tau,\xi)\bigr)$ into Eq.~\eqref{model}, where the phase of the plane wave background is $\theta_i=\mathrm{i}\bigl[b_i\tau+\bigl(\sigma(a_1^2+a_2^2)-\frac{1}{2}b_i^2\bigr)\xi\bigr]$, the perturbations are expressed in the standard Fourier eigenmode form $p_1=f_+e^{\mathrm{i}k(\tau+\Omega\xi)}+f_-^{*}e^{-\mathrm{i}k(\tau+\Omega^{*}\xi)}$ and $p_2=g_+e^{\mathrm{i}k(\tau+\Omega\xi)}+g_-^{*}e^{-\mathrm{i}k(\tau+\Omega^{*}\xi)}$, and $\Omega$ is the complex propagation eigenvalue.  Linearizing the resulting equations by neglecting higher-order nonlinear perturbation terms yields the linearized perturbation equations for $p_{1}$ and $p_{2}$: $\mathrm{i}\bigl(p_{i,\xi}+b_i p_{i,\tau}\bigr)+\frac{1}{2}p_{i,\tau\tau}+\sigma\sum_{l=1}^{2}a_l^{2}\bigl(p_{l}+p_{l}^{*}\bigr)=0$, $(i=1,2)$. These equations constitute the linearized BdG system for the perturbations \cite{BdG1,BdG2,BdG3}. By substituting the Fourier eigenmode form, we obtain the homogeneous BdG eigenvalue problem $\mathcal{K}\mathcal{P}=0$, where $\mathcal{K}={\rm diag}[(-\Omega-b_1-\frac{1}{2}k)k,(\Omega+b_1-\frac{1}{2}k)k,(-\Omega-b_2-\frac{1}{2}k)k,(\Omega+b_2-\frac{1}{2}k)k]+\sigma\mathcal{A},
\mathcal{A}=A[a_1^2,a_1^2,a_2^2,a_2^2], A=[1,1,1,1]^{T}, \mathcal{P}=(f_{+},f_{-},g_{+},g_{-})^T$. A nontrivial solution requires  $\mathrm{det}\mathcal{K}=0$, which yields the dispersion relation for linearized perturbations: $1+\sum_{i=1}^2\frac{\sigma a_i^2}{\left(\Omega+b_i\right)^2-\frac{1}{4}k^2}=0$. This yields a quartic equation in $\Omega$, giving four eigenvalues. For a given eigenvalue $\Omega$, solving $\mathcal{K}\mathcal{P}=0$ yields the corresponding perturbation eigenvector $\mathcal{P}=\varepsilon((k+2b_1+2 \Omega)^{-1},(k-2b_1-2 \Omega)^{-1},(k+2b_2+2 \Omega)^{-1},(k-2b_2-2 \Omega)^{-1})^{\mathrm{T}}$ \cite{Forest,Qin1}, where $\varepsilon$ is an arbitrarily small amplitude parameter. Among these solutions, the eigenmode associated with $\Omega$ with a negative imaginary part is physically unstable and grows exponentially along the propagation direction $\xi$. The MI gain is defined as $G=|\mathrm{Im}\Omega|$.

Without loss of generality, we consider $a_1=a_2=1$, $b_1=-b_2=\beta$. Then, the four eigenvalues $\Omega$ are $\pm\sqrt{\mu+\upsilon}$ and $\pm\sqrt{\mu-\upsilon}$, with $\mu=\beta^2+k^2/4-\sigma$ and $\upsilon=\sqrt{\beta^2(k^2-4\sigma)+\sigma^2}$. In the defocusing case ($\sigma=-1$), at most one pair of complex conjugate eigenvalues is admitted, i.e., only a single MI branch exists. In the focusing case ($\sigma=1$), which is of primary interest here, the four eigenvalues can become complex, forming two complex conjugate pairs under certain parameter conditions. Let $\Omega_1$ and $\Omega_2$ denote the eigenvalues with negative imaginary parts (the unstable ones), and $\Omega_1^*$, $\Omega_2^*$ their complex conjugates. The exact pairing between the four eigenvalue roots and the two conjugate pairs depends on the parameter regime: when $\upsilon^2<0$ the conjugate pairs are $\{\Omega_1, \Omega_1^*\}=\{-\sqrt{\mu+\upsilon},-\sqrt{\mu-\upsilon}\}$ and $\{\Omega_2,\Omega_2^*\}=\{\sqrt{\mu-\upsilon},\sqrt{\mu+\upsilon}\}$; when $\upsilon^2>0$ and $\mu+\upsilon<0$, they are $\{\Omega_1, \Omega_1^*\}=\{-\sqrt{\mu+\upsilon},\sqrt{\mu+\upsilon}\}$ and $\{\Omega_2, \Omega_2^*\}=\{-\sqrt{\mu-\upsilon},\sqrt{\mu-\upsilon}\}$.  However, the MI gains are universally given by $G_{1}=|\rm{Im}\Omega_1|$ and $G_{2}=|\mathrm{Im}\Omega_2|$. This yields two coexisting MI branches: MI$_{1}$ (gain $G_{1}$) and MI$_{2}$ (gain $G_{2}$).  Consequently, the two coexisting MI branches are described by distinct perturbation eigenvectors $\mathcal{P}_i$, where $i=1,2$ denote modes $\Omega_1,\Omega_2$, and $i=3,4$ their conjugates. By selecting the eigenvector associated with the desired branch in Eq.~\eqref{initial} \cite{Qin2}, one can favor its growth and ultimately realize controllable excitation of the corresponding vector AB pattern. Typical examples of the intensity evolution for the two MI branches are shown in Fig.~\ref{fig1}(a1,a2) ($\Omega_1=-1.14539-0.715487\mathrm{i}$, MI$_1$) and Fig.~\ref{fig1}(b1,b2) ($\Omega_2=1.14539-0.715487\mathrm{i}$, MI$_2$), confirming that this method enables controlled excitation of distinct vector AB patterns when two MI branches coexist on the same plane wave background.

Furthermore, by properly selecting the coefficient $\varepsilon$ within the perturbation eigenvector, we can further make numerical results highly consistent with the known exact vector AB solutions, including predicting identical spatiotemporal distributions. The exact AB solution of Eq.~\eqref{model} with period $2\pi/\alpha$ is given by
$\psi_{i,ana}=a_i e^{\theta_i}\frac{\chi_{1,1}^{(i)}e^{\omega_R}+\chi_{2,2}^{(i)}e^{-\omega_R}+\chi_{1,2}^{(i)}\mathcal{B}e^{{\rm i} \omega_I}+\chi_{2,1}^{(i)}\mathcal{B}^*e^{-{\rm i} \omega_I}}{2\cosh\left(\omega_R\right) +\mathcal{B}e^{{\rm i} \omega_I}+\mathcal{B}^*e^{-{\rm i} \omega_I}}$ ($i=1,2$)\cite{lingzhao2}, with \(\omega_I=\alpha\left[\tau+\left(\chi_{1R}+\frac {\alpha}{2}\right)\xi\right], \omega_R=\alpha\chi_{1I}\xi, \chi_1\equiv\chi_{1R}+\mathrm{i}\chi_{1I}\), \(\chi_2=\chi_1+\alpha\), \(\chi_{k,j}^{(i)}=\frac {\chi_k^*+b_i}{\chi_j+b_i}\), \(\mathcal{B}=\frac {2\chi_{1I}}{2\chi_{1I}-\rm i\alpha}\). Here, $\chi_1$ is determined by the AB governing equation $1+\sum_{i=1}^2\frac{\sigma a_i^2}{\left(\chi_1+b_i\right)\left(\chi_1+\alpha+b_i\right)}=0$.
The value of $\varepsilon$ is determined from dominant Fourier components of the exact solution under the early weakly modulated state. Specifically, we adopt the exact AB solution at an early instant of the evolution with extremely weak periodic modulation on the plane wave background. Instead of directly adopting the complete exact solution as initial data, we decompose it into dominant Fourier components and establish two-component initial fields as plane wave background superimposed with small linear perturbations, as expressed in Eq.~\eqref{initial}. The coefficient  $\varepsilon$ is then confirmed by matching perturbation eigenvectors to corresponding Fourier component amplitudes of the exact solution at this instant. This weakly modulated initial state seeds the growth of the AB in the simulations.  As a result, the numerically evolved solution remains in near-perfect agreement with the exact analytical solution, which is evidenced by the high fidelity between them. Here, fidelity is defined as
$F_i=\frac{|\int_{-L}^L\psi_{i,ana}^*\psi_{i,num}d\tau|^2}{\int_{-L}^L|\psi_{i,ana}|^2d\tau\int_{-L}^L|\psi_{i,num}|^2d\tau}$,
where the integrals are taken over the entire computational domain $[-L,L]$ with $L$ much larger than the breather width.
Therefore, this initial configuration not only realizes controllable AB excitation, but also further improves the matching degree between numerical simulations and exact solutions, achieving nearly 100\% fidelity, as displayed in Fig.~\ref{fig1}(a3) and (b3) for both MI branches.

The eigenvector-based initial condition Eq.~\eqref{initial} yielding this controllable high-fidelity excitation can also be equivalently written in an experimentally accessible amplitude-phase form as $\psi_i(\tau,0)=a_ie^{\mathrm{i}b_i\tau}\left[1+A_i(\tau)e^{\mathrm{i}\phi_i(\tau)}\right]$, where $A_i(\tau)$ and $\phi_i(\tau)$ are determined by the four complex Fourier coefficients $(f_+,f_-,g_+,g_-)$ fixed by the selected MI eigenvector. It thus directly determines the weak amplitude and phase modulations applied to the two continuous-wave pumps. The experimental feasibility of implementing such modulations has been established~\cite{Frisquet2014,Speirs2023,Kibler2015,Gelash2022}, while orthogonally polarized propagation in the Manakov regime has also been demonstrated~\cite{Frisquet2015,fiber4,fiber5,Liu2025}. For example, taking the physical modulation frequency as $50~\mathrm{GHz}$, the dimensionless parameters used in Fig.~\ref{fig1} correspond to a $100~\mathrm{GHz}$ pump separation and approximately $1.54~\mathrm{W}$ of background power per pump for representative standard-fiber parameters. These scales closely match those used in a recent two-component Manakov-breather experiment~\cite{Liu2025}, supporting the experimental feasibility of the proposed scheme.

\section{The Physical Mechanism of Controllable Akhmediev Breather Excitations}\label{sec3}

\begin{figure}[t!]
\centering		
\includegraphics[width=62mm]{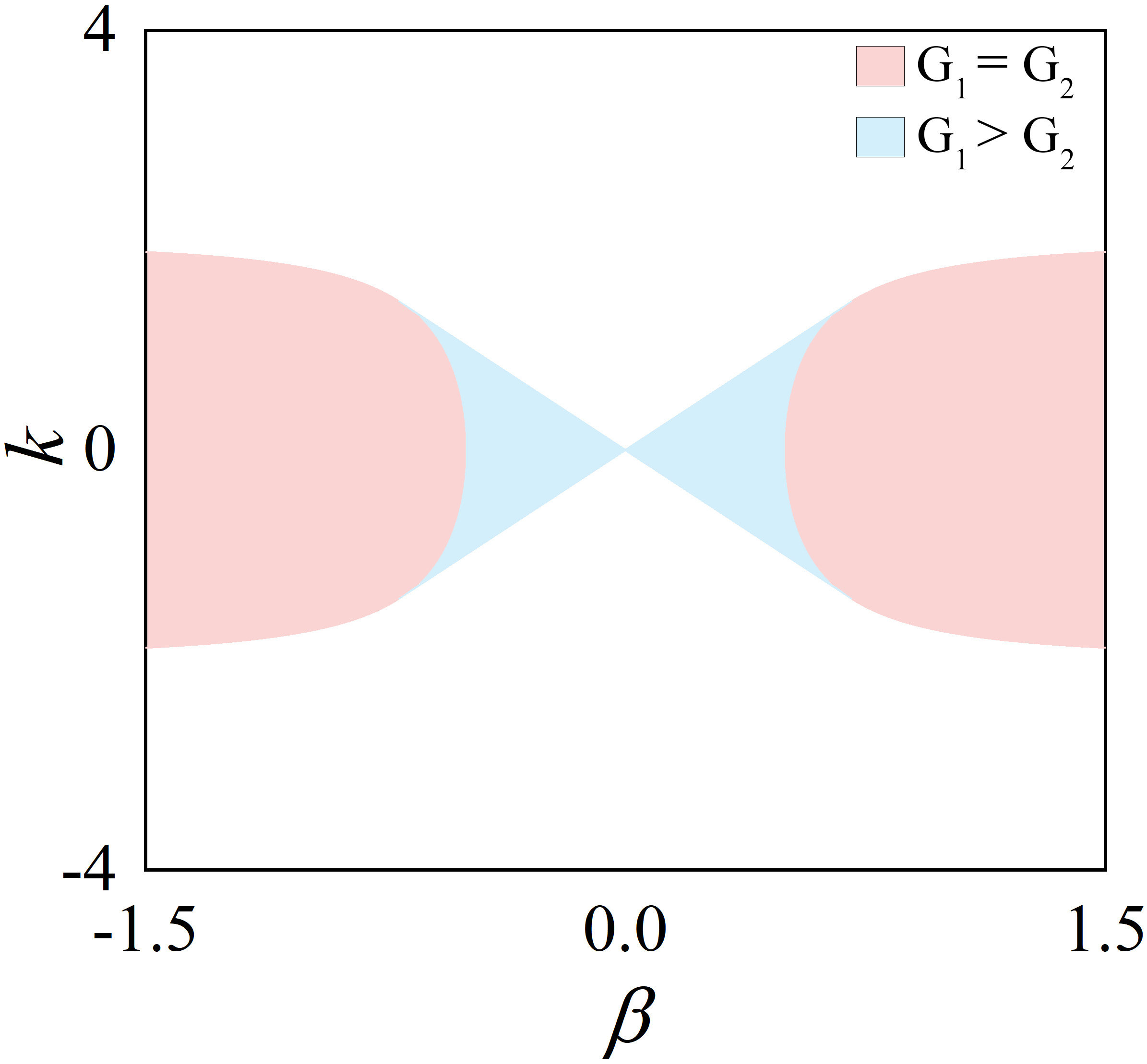}
\caption{Parameter regime for controllable excitation of vector ABs via our scheme, depicted in the $(\beta,k)$-plane. This region coincides with the coexistence regime of two MI branches in the focusing Manakov system. The pink region ($G_1=G_2$) corresponds to equal MI gains for the two branches, while the blue region ($G_1>G_2$) represents the case with unequal MI gains. The parameters are set as $a_1=a_2=1$, $b_1=-b_2=\beta$. }\label{fig2}
\end{figure}

Our analysis reveals that controllable excitations of vector ABs are widely feasible in the focusing Manakov system. We map the corresponding parameter regime on the $(\beta,k)$-plane in Fig.~\ref{fig2}. As this work concentrates on controllable excitation in the presence of multiple MI branches, we only present the parameter space where two MI branches coexist. Within this dual-branch regime, perfect excitation is stably realized in the pink region satisfying  $G_1=G_2$, where the two unstable eigenmodes possess identical growth rates. In the blue region with $G_1>G_2$, controllable excitation is still accessible by selectively prioritizing the MI$_1$ branch. Essentially, this controllability originates from the precise eigenvector selection encoded in the initial condition Eq.~\eqref{initial}, where the initial perturbation is aligned with the eigenvector associated with the targeted MI branch to achieve selective excitation.

To quantitatively characterize the dynamic evolution of perturbation eigenmodes, we adopt a standardized full BdG modal decomposition method. We first eliminate the background phase of the numerical wavefield and extract the Fourier components at $\pm k$ via Fourier analysis, constructing a $\xi$-dependent four-dimensional state vector $\mathcal{V}(\xi)$. Because the BdG operator is non-Hermitian, its right eigenvectors are generally nonorthogonal \cite{BdG2,NH2}. We therefore use the normalized right eigenvectors $\widetilde{\mathcal{P}}_i$ associated with $\{\Omega_1,\Omega_1^*,\Omega_2,\Omega_2^*\}$ and their dual left eigenvectors $\widetilde{\mathcal{L}}_i$, satisfying $\widetilde{\mathcal{L}}_i^\dagger\widetilde{\mathcal{P}}_j=\delta_{ij}$. The expansion coefficients of the normalized instantaneous state $\widetilde{\mathcal{V}}(\xi)$ are then $c_i(\xi)=\widetilde{\mathcal{L}}_i^\dagger\widetilde{\mathcal{V}}(\xi)$, and the normalized relative modal weights are defined by $w_i(\xi)=|c_i(\xi)|^2/\sum_{j=1}^{4}|c_j(\xi)|^2$, so that $0\leq w_i\leq1$ and $\sum_iw_i=1$. A value closer to $1$ indicates stronger dominance of this eigenmode. With the biorthogonal projection, a state exactly aligned with one right eigenvector has zero weight in all other modes. Therefore, for the four-dimensional perturbation vector $\widetilde{\mathcal{V}}(\xi)$, nonzero off-target weights quantify its deviation from the selected MI eigenvector within the fixed BdG basis. Beyond the linear regime, these quantities serve as diagnostics for the relative weights of this perturbation vector.

\begin{figure}[t!]
\centering		
\includegraphics[width=85mm]{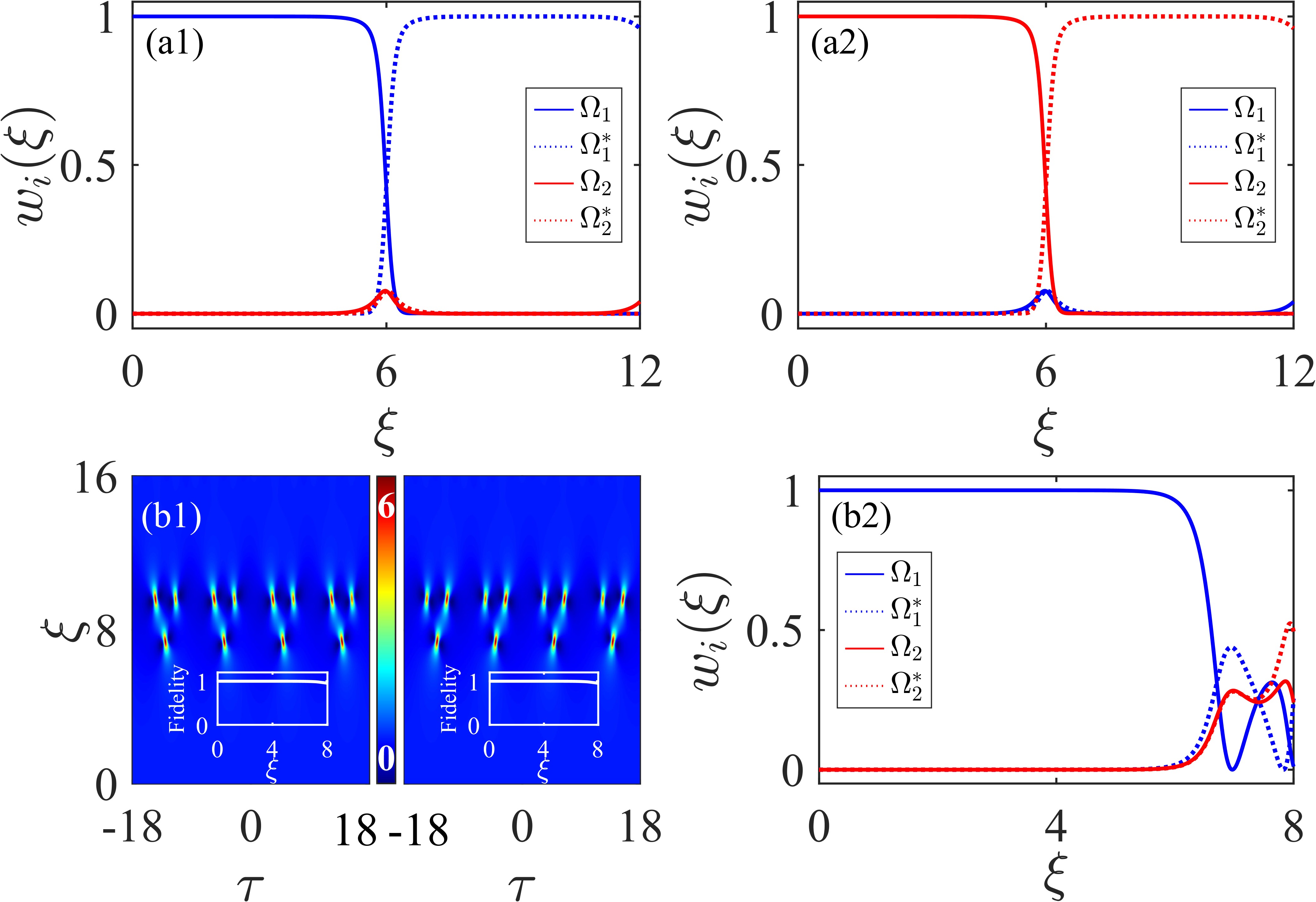}
\caption{(a1)-(a2): Normalized relative weight of each BdG eigenmode, within a clean AB formation (shown in Fig.~\ref{fig1}) for the MI$_1$ and MI$_2$ branches, respectively. (b1): Intensity evolution of a nearly perfect vector AB excitation when selecting the MI$_1$ branch, with $G_1$ significantly larger than $G_2$. Inset: Fidelity between the numerical AB (before the emergence of frequency doubling) and the exact AB solution; the minimum fidelity is approximately 95\% as subsequent frequency doubling distorts the waveform. (b2): Corresponding normalized relative weight of each BdG eigenmode for the case shown in (b1). The parameters for (b1) and (b2) are: $a_1=1,a_2=1,b_1=0.4,b_2=-b_1,k=0.7$ and $\varepsilon=-0.0017266 + 0.0005145\mathrm{i}$.}\label{fig3}
\end{figure}

The modal decomposition is then applied to the cases shown in Figs.~\ref{fig1}(a1,a2) and~\ref{fig1}(b1,b2). The resulting evolutions of the relative modal weights are plotted in panels \ref{fig3}(a1) and \ref{fig3}(a2) of Fig.~\ref{fig3}, respectively. It is seen that the targeted unstable eigenmode associated with $\Omega_1$ (or $\Omega_2$) dominates the breather generation, and a clear reversal of dominance from this growing mode to its conjugate decaying mode occurs once the AB structure is formed. The other off-target modes maintain low weights throughout the dynamical evolution. This confirms that the dominant mode during the breather formation stage is directly selected by the initial condition through the eigenvector structure of the desired MI branch, and its subsequent dynamical dominance is then quantified by the instantaneous relative weights, with a larger value signifying a greater influence on the state evolution.

The above numerical modal decomposition intuitively presents the macroscopic modal evolution characteristics during vector AB excitation. On this basis, we further reveal the physical mechanism behind selective excitation by analyzing the structure of the initial perturbation and the resulting modal competition. The initial condition Eq.~\eqref{initial} is exactly the $\xi=0$ form of the Fourier perturbations $p_1$ and $p_2$. The four coefficients $(f_+,f_-,g_+,g_-)$ are completely determined, up to an overall factor, by the right eigenvector $\mathcal{P}$ of the target MI branch, so that the initial perturbation is aligned with the corresponding eigenmode. Therefore, the biorthogonal decomposition assigns the entire initial weight to the selected mode and zero initial weight to all off-target modes. During the early linear stage, the selected unstable mode grows exponentially and dominates the modulation of the uniform background. This initial modal dominance allows the target branch to dictate the formative stage of the AB pattern and thereby prescribe its structural type. Although the subsequent nonlinear evolution induces a dominance reversal from the growing mode to its conjugate decaying mode within the selected branch, as is characteristic of AB formation, this reversal does not alter the branch identity of the breather. Consequently, the eigenvector chosen in the initial condition determines which type of vector AB is excited.

This dominance reversal reflects the homoclinic evolution of the AB within the selected MI branch. In the BdG spectrum, each MI branch consists of a pair of complex-conjugate eigenvalues, with their respective eigenvectors. The initial condition Eq.~\eqref{initial} is constructed solely from the eigenvector $\mathcal{P}$ of the chosen unstable eigenvalue $\Omega$, and its Fourier form only seeds the $\Omega$-mode at the initial plane. During the early linear growth stage, the $\Omega$-mode grows exponentially, and dominates the formation of the localized AB structure, drawing energy from the background into the breather \cite{RW4,Forest,Baronio1,Baronio2,lingzhao2,zhaoMI}.
Once the breather is fully formed, the relative weight of the conjugate decaying $\Omega^*$-mode within the same MI branch increases and becomes dominant, marking the nonlinear return stage during which energy flows from the breather back to the plane wave background. The two modes thus dominate the growth and decay stages, respectively, as the system completes one Fermi-Pasta-Ulam recurrence cycle characteristic of integrable dynamics \cite{FPU1,FPU2,FPU3,FPU4}. As a result, the generated vector ABs exhibit high purity and controllability.

In contrast to the modes associated with the target MI branch, the off-target eigenmodes $\Omega_2$ and $\Omega_2^*$ (or $\Omega_1$ and $\Omega_1^*$) have zero initial weights but can develop small nonzero relative weights during the subsequent evolution, particularly when nonlinear effects become significant. According to the biorthogonal decomposition introduced above, these nonzero weights quantify the small components of the instantaneous perturbation vector along the eigenmodes associated with  the other MI branch. In the clean-excitation cases, the off-target weights remain small and do not change the dominance of the selected MI branch. These small off-target weights should be distinguished from the dominant intra-branch reversal discussed above.

The above analysis establishes that the AB pattern is determined by the target MI branch, which is selectively seeded through its eigenvector and dominates the early linear stage. This eigenvector-controlled selection is most clearly seen in the pink gain-balanced region of Fig.~\ref{fig2}.
A more stringent test is whether the same eigenvector-based control remains robust under extreme gain imbalance, when one branch possesses a much larger growth rate than the other. We first target the branch with the stronger instability. In the blue region of Fig.~\ref{fig2}, where $G_1$ is significantly larger than $G_2$, high-fidelity controllable excitation of vector ABs can indeed be achieved by targeting the stronger MI$_1$ branch. For example, we set $a_1=a_2=1$, $\beta=0.4$, $\alpha=0.7$, $k=\alpha$. These parameters yield eigenvalues $\Omega_1=-1.17457\mathrm{i}$ and $\Omega_2=-0.235334\mathrm{i}$. For these settings, the exact solution $\psi_{i,ana}$ analytically admits two possible AB patterns: an eye-shaped one, associated with the MI$_1$ branch, and a four-petaled one, associated with the MI$_2$ branch. By constructing the initial condition in Eq.~\eqref{initial} using the eigenvector $\mathcal{P}_1$, the numerical intensity evolution in Fig.~\ref{fig3}(b1) confirms the generation of an eye-shaped AB pattern, and the first fundamental AB that emerges closely matches the exact solution. Because the perturbation frequency lies in the high-order MI region \cite{HMI}, frequency doubling emerges after the fundamental AB. Despite this, the fundamental AB maintains high fidelity with the exact solution, with the minimum fidelity reaching as high as 95\% (inset of Fig.~\ref{fig3}(b1)); the slight deviation from 100\% is associated with the subsequent frequency doubling. The corresponding normalized relative weights of the eigenmodes are shown in Fig.~\ref{fig3}(b2). As can be seen, the target $\Omega_1$ mode dominates the early growth of the fundamental AB. Finite weights of the other eigenmodes appear near maximum compression as nonlinear effects become significant, and their subsequent evolution is further affected by frequency doubling. This result confirms that eigenvector-based seeding can generate the desired AB pattern when the target MI branch has a substantial gain advantage over the coexisting branch.

\begin{figure}[t!]
\centering		
\includegraphics[width=85mm]{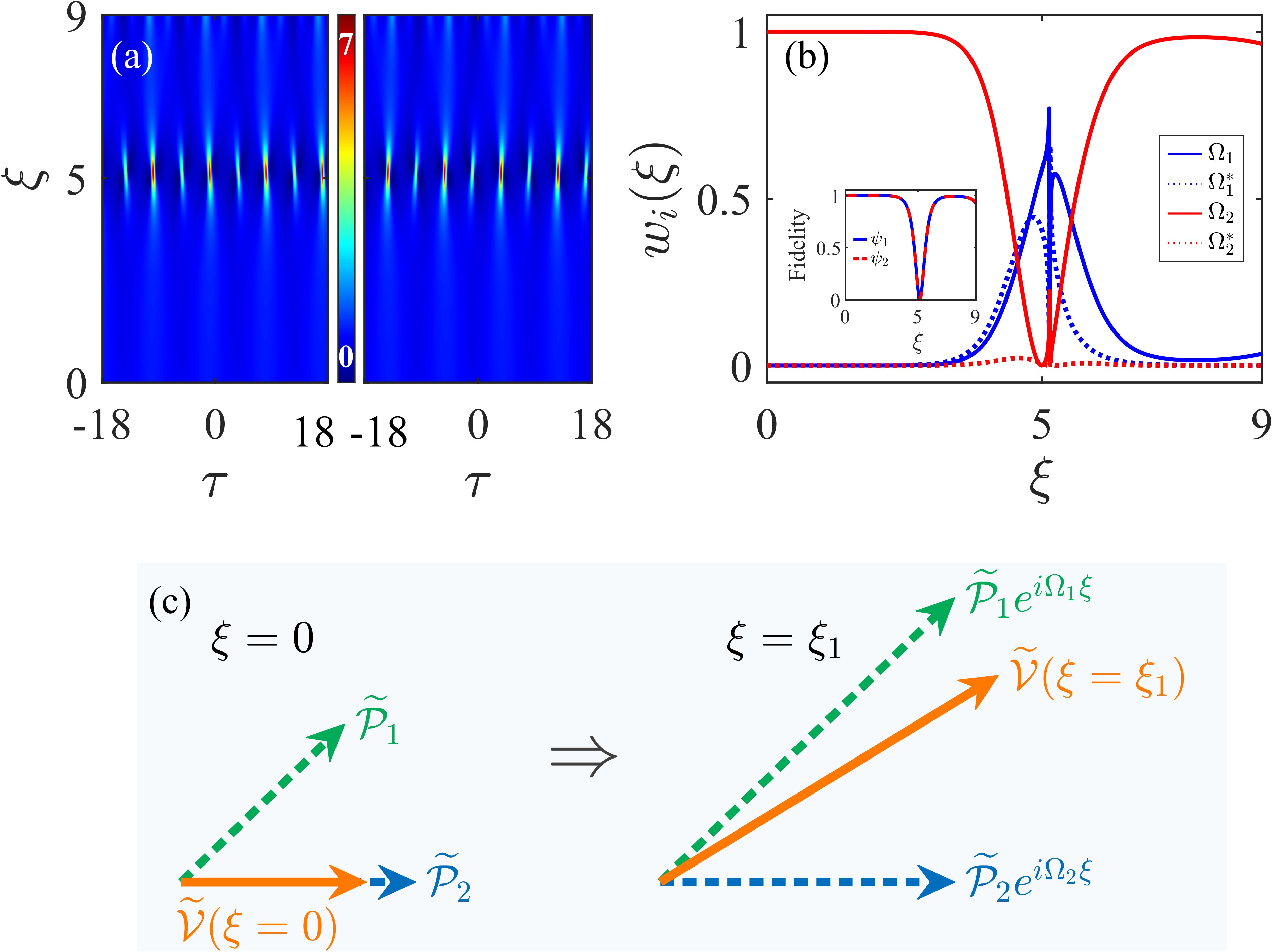}
\caption{Failed excitation of the desired vector AB when targeting the MI$_2$-branch under the large gain imbalance condition $G_1 > G_2$.
(a): Intensity evolution of the two components simulated from the initial condition with  $\Omega = -0.235334\mathrm{i}$ with $\varepsilon = -0.009528 + 0.014142\mathrm{i}$; with all other parameters identical to those used in Fig.~\ref{fig3}(b1). (b): Evolution of normalized relative modal weights obtained via full BdG eigenbasis projection. The inset shows the fidelity between the numerically evolved field and the exact vector-AB solution. (c): Schematic of the normalized instantaneous state $\widetilde{\mathcal{V}}(\xi)$ (orange) relative to the two linear BdG eigenmode solutions $\widetilde{\mathcal{P}}_i e^{\mathrm{i}\Omega_i\xi}$ associated with the MI$_1$ and MI$_2$ branches (green and blue); their conjugate modes are omitted here for simplicity. Initially, $\widetilde{\mathcal{V}}(0)$ is aligned with $\widetilde{\mathcal{P}}_2$. During the nonlinear evolution, it develops a finite projection onto $\widetilde{\mathcal{P}}_1$, and the gain disparity $G_1>G_2$ makes the MI$_1$ contribution dominant.}\label{fig4}
\end{figure}

In stark contrast, targeting the weaker $\mathrm{MI_2}$ branch fails to generate the expected four-petaled vector AB solution, as demonstrated by the spatiotemporal intensity evolution in Fig.~\ref{fig4}(a). The initial condition constructed from $\mathcal{P}_2$ places the entire initial normalized modal weight on the targeted $\Omega_2$ eigenmode (red curve, Fig.~\ref{fig4}(b)). As nonlinear effects become significant, however, the competing $\Omega_1$ mode develops a finite normalized weight. Owing to the large gain imbalance $G_1>G_2$, its modal contribution rapidly increases and eventually dominates the evolution. This modal competition is also illustrated in the decomposition sketch of Fig.~\ref{fig4}(c). At $\xi=0$, the state vector $\widetilde{\mathcal{V}}(0)$ is aligned with the targeted eigenmode $\widetilde{\mathcal{P}}_2$ of the MI$_2$ branch. The linear BdG eigenmode solutions are characterized by their respective phase factors $e^{\mathrm{i}\Omega_i\xi}$. Because the MI$_1$ branch has a much larger gain, once a finite projection onto $\widetilde{\mathcal{P}}_1$ develops during the subsequent nonlinear evolution, the associated modal contribution can rapidly become dominant, tilting the instantaneous state vector $\widetilde{\mathcal{V}}(\xi)$ toward the MI$_1$ eigenmode and breaking the original alignment with $\widetilde{\mathcal{P}}_2$. Consequently, significant crosstalk develops between the coexisting MI branches. The resulting structure does not correspond to a pure MI$_2$- or MI$_1$-type single-branch vector AB. It features a doubled breathing frequency compared with the theoretical single-branch solution, as also shown in the inset of Fig.~\ref{fig4}(b), where the fidelity decreases sharply as nonlinear effects become significant.

Therefore, successful eigenvector-controlled mode selection requires a weak initial perturbation accurately described by the corresponding MI eigenvector, together with comparable gains between the coexisting MI branches or a sufficient gain advantage for the target branch. Under these conditions, that branch can remain dominant throughout the formation of the target AB pattern, as demonstrated in Fig.~\ref{fig3}. When the target branch has a substantially lower gain, the higher-gain branch can overtake it and cause cross-talk before the intended AB forms, as shown in Fig.~\ref{fig4}. These results show that controllable excitation of vector AB patterns requires the target MI branch to maintain its dominance during competition between the coexisting branches.

\section{Conclusion and Discussion}\label{sec4}

We have demonstrated controllable high-fidelity excitation of vector ABs in the focusing Manakov system using a simple eigenvector-based initial condition. The initial state consists of a plane wave background plus weak Fourier modes whose amplitudes and phases are determined by the perturbation eigenvector of a selected MI branch. Numerical simulations show that this scheme can generate the target vector AB with near-100\% fidelity to the exact solution in gain-balanced regimes or when the target branch has a sufficient gain advantage. The underlying mechanism is eigenvector-controlled mode selection within the non-Hermitian BdG eigenstructure. Matching the initial perturbation to the right eigenvector of the target branch concentrates the initial modal weight in the corresponding unstable mode, while the relative MI gains determine whether this modal dominance persists throughout AB formation. However, when the target branch is significantly weaker, the higher-gain branch can overtake it during the nonlinear evolution, leading to cross-talk and preventing the formation of the intended single-branch AB. Selective excitation of such a weak branch remains challenging and may require adiabatic parameter ramping to prevent the stronger branch from rapidly becoming dominant or additional degrees of freedom to control interbranch competition. More generally, they show that pattern selection in multi-component systems with coexisting instabilities is governed by the interplay between the initial eigenvector choice and the relative MI gain rates.

These results provide a route to overcoming the difficulty of exciting clean vector AB patterns from simple initial perturbations when multiple MI branches coexist. More broadly, this eigenvector-based selection principle could provide a useful mode-selection strategy for systems with multiple instability channels, such as nonlinear optics \cite{fiber4,fiber5,Liu2025}, Bose-Einstein condensates \cite{BEC3,rwBEC}, and hydrodynamic systems \cite{Onorato}. It may also offer an approach to controlling the initialization of complex vector frequency-comb states in driven-dissipative Kerr systems \cite{Xu2021,Xu2022,Lucas2025Faticon}, with potential relevance to spatially and polarization-multiplexed Kerr combs in modern microresonator physics \cite{Lucas2018,Bao2019,Geng2022}.

\section*{ACKNOWLEDGMENTS}

We are grateful to Haiyang Yu, Ning Mao, and Tao Jiang for their helpful discussions. Y.-H. Q. is supported by the National Natural Science Foundation of China (Grant No. 12405004), the Natural Science Foundation of Xinjiang Uygur Autonomous Region Project (Grant No. 2024D01C232), the Scientific Research Projects Funded by the Basic Research Business Expenses of Autonomous Region Universities (Grant No. XJEDU2024P011), the ``Tianchi Talent" Introduction Plan in Xinjiang Uygur Autonomous Region, and the Tianshan Innovation Team Program of Xinjiang Uygur Autonomous Region (Grant No. 2026D14003). L.-C. Z. is supported by the National Natural Science Foundation of China (Contracts No. 12375005 and No. 12235007), and the Major Basic Research Program of Natural Science of Shaanxi Province (Grant No. 2018KJXX-094).

\end{document}